\documentclass[aps,twocolumn,superscriptaddress,amsmath,amssymb]{revtex4-1}
\usepackage{graphicx}  % needed for figures 
\usepackage{dcolumn}   % needed for some tables
\usepackage{bm}        % for math 
\usepackage{verbatim}   % for math
\usepackage{color} 
\usepackage{physics}
\usepackage{amsmath}
\UseRawInputEncoding

\begin{document}
\title{Non-collinear first-principles studies of the spin-electric coupling in frustrated triangular molecular magnets}

\author{M. F. Islam}
\affiliation{Department of Physics, Central Michigan University, Mount Pleasant, MI-48859, USA}
\affiliation{Department of Physics, The University of Texas at El Paso, TX-79968, USA}
\author{Kushantha P. K. Withanage}
\affiliation{Department of Physics, The University of Texas at El Paso, TX-79968, USA}
\author{C. M. Canali}
\affiliation{Department of Physics and Electrical Engineering,Linnaeus University, SE-39182 Kalmar, Sweden}
\author{Mark R. Pederson}
\affiliation{Department of Physics, The University of Texas at El Paso, TX-79968, USA}

\date{\today}

\begin{abstract}
Frustrated triangular molecular magnets (MMs) with anti-ferromagnetic ground states (GS) are an important class of magnetic systems with potential applications in quantum information processing. The two-fold degenerate GS of these molecules, characterized by spin chirality, can be utilized to encode qubits for quantum computing. Furthermore, because of the lack of inversion symmetry in these molecules, an electric field couples directly states of opposite chirality, allowing a very efficient and fast control of the qubits. In this work we present a theoretical method to calculate the spin-electric coupling for triangular MMs with effective {\it local} spins $s$ larger than 1/2, which is amenable to a first-principles implementation based on density functional theory (DFT). In contrast to MMs where the net magnetization at the magnetic atoms is $\mu_{\rm B}/2$ ($\mu_{\rm B} $ is the Bohr magneton), the DFT treatment of frustrated triangular MMs with larger local magnetizations requires a fully non-collinear approach, which we have implemented in the NRLMOL DFT code.
As an example, we have used these methods to evaluate the spin-electric coupling for a spin $s = 5/2$ $\{\mathrm{Fe_3}\}$ triangular MM, where this effect has been observed experimentally for the first time quite recently. Our theoretical and computational methods will help elucidate and further guide ongoing experimental work in the field of quantum molecular spintronics.
\end{abstract}

\maketitle

%%%%%%%%%%%%%%%%%%%%%%%%%%%%%%%%%%%%%%%%%%%%%%%%%%%%%%%%%%%%%%%%%%%%
%                 section     Introduction
%%%%%%%%%%%%%%%%%%%%%%%%%%%%%%%%%%%%%%%%%%%%%%%%%%%%%%%%%%%%%%%%%%%%

\section{Introduction}
\label{Intro}

Since the synthesis of $\{\mathrm{Mn_{12}}\}$ MM\cite{Sessoli1993} and the subsequent investigation of its potential as a magnetic storage device and its remarkable macroscopic quantum effects\cite{Leuenberger2001}, many different types of MMs are presently being studied for possible applications in quantum spintronics and quantum information processing\cite{Barra1996,Christou2000,Nossa2013,Tesi2016,Gabbi2019,Gaita2019,Hooshmand2021,Lunghi2022,Fursina2023}. The ability to manipulate the spin states of a magnetic molecule efficiently is a crucial aspect for the successful realization of molecular quantum spintronics devices. The spin states in a MM can be manipulated naturally by an external magnetic field. However, focusing a magnetic field on a nanoscale region is hard, expensive and inefficient. Alternatively, for a system with a strong spin-orbit coupling, an electric field can be used to control the spin states via a modification of the electronic orbitals. However, this method is not efficient either as the spin-orbit energy scales with volume of the system and, therefore, is weak for molecules. 

Therefore, a molecular system that does not require strong spin-orbit coupling but for which the spin states can be manipulated by an external electric field is highly desirable for the practical implementation of MM devices. Frustrated triangular MMs are very promising candidates of this system. The anti-ferromagnetic (AFM) GS manifold of these molecules is four-fold degenerate\footnote{This is strictly true only in the absence of spin-orbit coupling. A DM-induced spin-orbit coupling slightly lifts the degeneracy\cite{Trif2008, Nossa2012}} and comprises states characterized by a total spin projection quantum number $S_z = \pm 1/2$ as well as by a two-fold spin chirality eigenvalue $C_z = \pm 1$, which can be used to encode qubits for quantum information processing. The mechanism of spin-electric coupling in frustrated triangular molecules has been proposed independently by two groups of researchers. Using a Hubbard model at half filling, Bulaevskii {\it et. el.}\cite{Bulaevskii2008} have shown that for a frustrated equilateral triangular system, the average charge density at a given site depends on the spin configuration and it is not uniform for the antiferromagnetic spin configurations that contribute to the GS manifold. This leads to a net electric dipole moment in the triangle, which, in turn, can couple with an external electric field. Soon after, Trif {\it et. el.}\cite{Trif2008}, using a group theoretical approach, have shown that in triangular molecules with $\mathrm{D_{3h}}$ symmetry, the electric field can couple states of the GS manifold with opposite spin chirality. While these methods explain the basic physical mechanism of the phenomenon, they are not able to elucidate the microscopic details and especially to evaluate the strength of the spin-electric coupling in realistic triangular MMs with complicated crustal and electronic structure. In order to resolve the issue, we have developed a method that allows us to calculate the coupling strengths of realistic spin $s= 1/2$ MMs using DFT methods and have applied them to calculate coupling strengths in $\{\mathrm{Cu_3}\}$\cite{Islam2010} and $\{\mathrm{V_3}\}$\cite{Nossa2023} triangular spin $s= 1/2$ MMs.  
Following early theoretical work, a few other theoretical papers were published on the subject over the years, mainly reviewing the basic principles of the spin-electric coupling\cite{Khomskii2010,Khomskii2012} or extending the original analysis to more general settings\cite{Trif2010,Nossa2012, Nossa2023}. However, the first experimental evidence of spin-electric coupling in frustrated molecules had to wait a decade after the original theoretical predictions, when Boudalis {\it et. el.}\cite{Boudalis2018} published a study carried out in a spin $s = 5/2$ triangular MM, namely in the $\mathrm{[Fe_3O(O_2CPh)_6(Py)_3]ClO_4.Py}$ molecular complex, henceforth called here $\{\mathrm{Fe_3}\}$ MM.  By employing electron paramagnetic resonance (EPR) techniques, they observed that the intensity of the absorption spectrum increases when a static electric field is applied parallel to the triangular plane of the molecule, a conclusive signature of spin-electric coupling in this molecule. More recently, a direct observation of the spin-electric effect in the same $\{\mathrm{Fe_3}\}$ SMM has also been reported by Lewkowitz {\it et. el.}\cite{Lewkowitz2023}. The spin-electric coupling has also been observed in a different $\{\mathrm{Cu_3}\}$\cite{Liu2019} and $\{\mathrm{Co_3}\}$\cite{Kintzel2021} triangular molecular complexes. This series of successful experimental observations of the spin-electric coupling in different types of triangular MMs has strongly renewed interest in this class of MMs.

As mentioned above, all these first experimental observations deal with spin $s= 5/2$ triangular MMs. However, most of the theoretical analysis carried out over the past decade deals with spin $s= 1/2$ triangular MMs. In particular, no theoretical and computational method have been developed yet to calculate the spin-electric coupling for $s = 5/2$ MM {\footnote{In the important Ref.~\cite{Trif2010}, Trif {\it et al.} used a group-theory analysis to show that the this coupling is allowed also spin in $s=3/2$ triangular MMs. However, the considerably more complex spin $s=5/2$ was not investigated}}. Importantly, the use of first-principles methods to evaluate the strength of this coupling for $s > 1/2$ triangular MMs requires a non-trivial extension of the DFT approach that we have developed for the $s=1/2$ case\cite{Islam2010,Nossa2012,Johnson2019,Nossa2023}, since in this case a fully non-collinear (NCL) approach is unavoidable in order to handle frustrated spin configurations.{\footnote{The spin $s=/2$ AFM triangular case is special in that, although frustrated, the quantum eigenstate at any magnetic site of the local spin projection along an arbitrary direction (a spin coherent state corresponding to a classical spin orientation along that direction) can always be written as convenient linear combination of quantum spin projection $s_z \pm 1/2$ states, where $z$ is quantization axis common to all three magnetic sites. Therefore, the non-collinearity in this case is in fact trivial.}}  

In this work we have developed such a NCL first-principles approach and implemented it in the NRLMOL DFT code\cite{Pederson1990,Jackson1990,Porezag1999}. Although our method is valid in general and applicable to any spin $s =(2N +1)/2, N \geq 0$ triangular AFM MM, we have applied it only to the most complex s= 5/2 case, which is the one relevant for ongoing experiments on real MMs. The theoretical method is described in Sec.~\ref{SEcoupling}. Details of the implementation in the DFT code are discussed in Sec.~\ref{ncm}. The results of the calculations of the spin-electric coupling for spin  $s= 5/2\;$ $\{\mathrm{Fe_3}\}$ MM are discussed in Sec.~\ref{Results}.    

\section{Theoretical model: Spin-electric coupling in spin $s= (2N+ 1)/2$ triangular magnets}
\label{SEcoupling}

In this work we have developed a general formula for the spin-electric coupling in the AFM ground state of any frustrated triangular SMM with half-integer spin. By utilizing this approach, the strength of the coupling can be calculated suitably using DFT methods. To study the magnetic properties of any frustrated triangular SMM, we first solve the Heisenberg model in the spin manifold of the triangular spin system,
\begin{equation}
    \mathcal{H}=J\sum_{ij}^3 S_i\cdot S_j, \hspace{0.5 cm}i,j > 0. 
\label{H_ex}
\end{equation}
Here $S_i, S_j = (2N + 1)/2$, where N = 0, 1, 2, 3 $\cdots$. The GS of the Hamiltonian is two doublets with total spin $S_T=1/2$: one doublet is associated with spin projection $S_z = +1/2$ and the other one is for $S_z = -1/2$. Since we are interested in the coupling in the GS of these systems, for this work, we focus only on the spin-1/2 subspace of the Heisenberg model. While the spin-manifold of three (2N + 1)/2 spins have a total of $(2N + 2)^3$ spin configurations, only $3(N+1)^2$ states form the bases for each of the spin projections $S_z = \pm 1/2$. It should be noted that out of $3(N+1)^2$ states, only $(N+1)^2$ states are independent. For each of these independent states, two more states are related by $C_3$ symmetry of the molecule.

The solution to Eq.~\ref{H_ex} are $\it {real}$, which are not manifestly chiral. Therefore, to construct the chiral form of the GS we consider only the GS manifold, we note that the scalar chiral operator for the three-spin system is defined by,
\begin{equation}
    C_z=\frac{4}{\sqrt{3}} S_1\cdot S_2 \cross S_3
\label{cz}
\end{equation}
with eigenvalues $\pm 1$. Since the chiral operator, $C_z$, commutes with the spin Hamiltonian in Eq.~\ref{H_ex}, they share common eigenstates. We can obtain the chiral states by diagonalizing the chiral operator on the basis of real GSs\cite{Nossa2023}, but here we have adopted an alternative approach. We consider the chiral operator as a small perturbation to the Hamiltonian.
\begin{equation}
    \mathcal{H}=J\sum_{ij}^3 S_i\cdot S_j + \lambda C_z, \hspace{0.5 cm}i,j > 0
\label{H_chi}
\end{equation}
Here $\lambda$ is a very small number, say $10^{-6}$. As in the case of Hamiltonian Eq.~\ref{H_ex}, GS is two-fold degenerate for each of the spin projections, $S_z=\pm 1/2$ but the amplitude of the spin bases are now complex. One of the GSs associated with $S_z=+1/2$ is obtained as 
\begin{equation}
\Psi^1_\mathrm{GS}(S_z=+1/2) = \sum_{i=1}^{(N+1)^2}(a_{i1}\ket{B_{i1}} + a_{i2}\ket{B_{i2}} + a_{i3}\ket{B_{i3}}) 
\label{gs}   
\end{equation}
where $B_{ij}$'s (i = 1, 2, $\cdots$; j = 1, 2, 3) are basis states in the GS-manifold and the corresponding amplitudes are $a_{ij}$. All the $3(N+1)^2$ basis states in the GS are grouped into $(N+1)^2$ subgroups. The three bases in each subgroup are related by $C_3$ symmetry. 

To recast the GS Eq.~\ref{gs} into the chiral form, we take the first amplitude $a_{i1}$ from each subgroup as a common factor. Then the resulting GS takes the form,
\begin{eqnarray}
\Psi^1_\mathrm{GS}(S_z=+1/2, \chi=+1) =  \nonumber \\  
= \sum_{i=1}^{(N+1)^2}a_{i1}(\ket{B_{i1}} + \frac{a_{i2}}{a_{i1}}\ket{B_{i2}} + \frac{a_{i3}}{a_{i1}}\ket{B_{i3}}) \nonumber \\
= \sum_{i=1}^{(N+1)^2} a_{i1}(\ket{B_{i1}} + \omega\ket{B_{i2}} + \omega^2\ket{B_{i3}})
%= a_{11}\Big(\ket{\frac{-5}{2} \frac{1}{2} \frac{5}{2}} + \omega\ket{\frac{5}{2} \frac{-5}{2} \frac{1}{2}} + \omega^2\ket{\frac{5}{2} \frac{-5}{2} \frac{1}{2}}\Big) \nonumber \\ 
% + a_{21}\Big(\ket{\frac{-5}{2} \frac{3}{2} \frac{3}{2}} + \omega\ket{\frac{3}{2} \frac{-5}{2} \frac{3}{2}} + \omega^2\ket{\frac{3}{2} \frac{3}{2} \frac{-5}{2}}\Big) &\nonumber \\ 
%+ \cdots \cdots \nonumber \\ 
%+ a_{91}\Big(\ket{\frac{-1}{2} \frac{1}{2} \frac{1}{2}} + \omega\ket{\frac{1}{2} \frac{-1}{2} \frac{1}{2}} + \omega^2\ket{\frac{1}{2} \frac{1}{2} \frac{-1}{2}}\Big) 
\label{cgs1}   
\end{eqnarray}
where, $\omega=e^{i2\pi/3}$. The state is also the eigenstate of the chiral operator, $C_z$ with eigenvalue +1. Similarly, we can obtain the chiral form of the second degenerate GS with opposite chirality and the same spin-projection from the second GS of the Hamiltonian Eq.~\ref{H_chi}.
\begin{eqnarray}
\Psi^2_\mathrm{GS}(S_z=1/2, \chi=-1) = \nonumber \\ 
= \sum_{i=1}^{(N+1)^2}(b_{i1}\ket{B_{i1}} + b_{i2}\ket{B_{i2}} + b_{i3}\ket{B_{i3}}) \nonumber \\
=\sum_{i=1}^{(N+1)^2} b_{i1}(\ket{B_{i1}} + \omega^2\ket{B_{i2}} + \omega\ket{B_{i3}}) 
\label{cgs2}   
\end{eqnarray}
We note that the magnitude of $b_{i1}$s are the same as corresponding $a_{i1}$s but has opposite phase, confirming the orthogonality of the two degenerate chiral GSs. 

\subsection*{Spin-electric coupling in triangular MMs}

\begin{figure}[ht]
\centering{\includegraphics[width=0.45\textwidth]{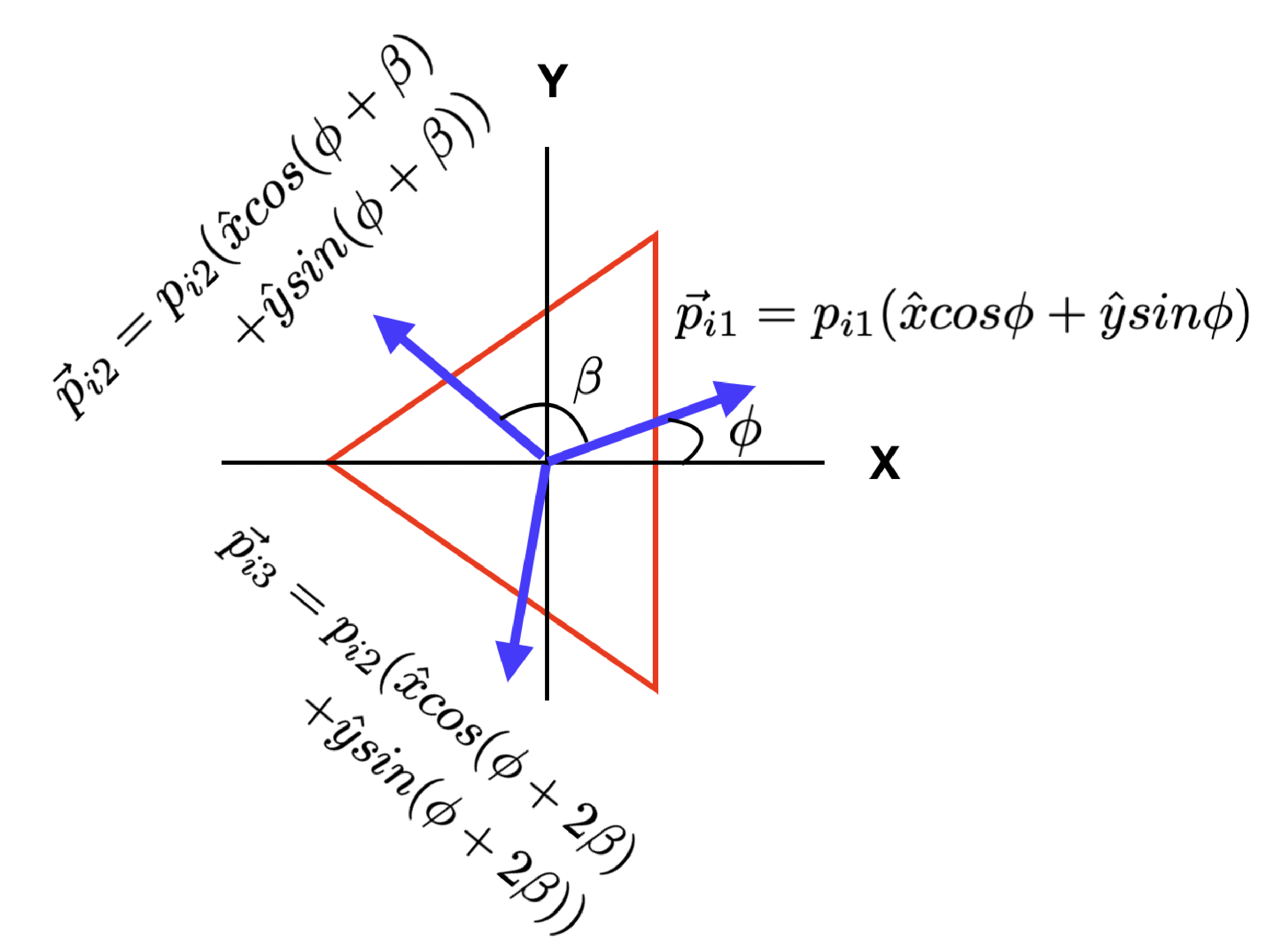}}
\caption{Orientation of the dipole moments of the three spin configurations related by $C_3$ symmetry.}
\label{dip}
\end{figure}

The spin-electric coupling between the two GSs of opposite chirality and the same spin-projection is the matrix element of the dipole operator in these two states and can be expressed as

\begin{equation}
\bra{\Psi^1_\mathrm{GS}}|e\Vec{E}\cdot\Vec{r}|\ket{\Psi^2_\mathrm{GS}} = e\Vec{E}\cdot\bra{\Psi^1_\mathrm{GS}}|\Vec{r}|\ket{\Psi^2_\mathrm{GS}} = e\Vec{E}\cdot\Vec{d}.
\label{se}   
\end{equation}
where $\vec{E}$ is the applied electric field. Substituting Eq.~\ref{cgs1} and \ref{cgs2} and using orthogonality of the spin states, we obtain 
\begin{eqnarray}
\Vec{d} &=& \sum_{i=1}^{(N+1)^2} a_{i1}^*b_{i1}(\bra{B_{i1}}\Vec{r}\ket{B_{i1}} + \omega\bra{B_{i2}}\Vec{r}\ket{B_{i2}} \nonumber \\ 
  &+& \omega^2\bra{B_{i3}}\Vec{r}\ket{B_{i3}}) \nonumber \\ 
  &=& \sum_{i=1}^{(N+1)^2} a_{i1}^*b_{i1}(\Vec{p}_{i1}+\omega\Vec{p}_{i2}+\omega^2\Vec{p}_{i3}) \nonumber \\
  &=& \sum_{i=1}^{(N+1)^2} a_{i1}^*b_{i1}\Vec{p}_i.
\label{SE1}   
\end{eqnarray}
Here, the sum is over the $(N+1)^2$ inequivalent basis states (the states that are not related by $C_3$ symmetry i.e. the states in the first column of Table~\ref{tab_basis}). $\Vec{p}_{i1}$, $\Vec{p}_{i2}$, and $\Vec{p}_{i3}$ are the dipole moments of the three spin configurations (see  Fig.~\ref{dip}) and have the same magnitude due to the $C_3$ symmetry. Furthermore, the $\sigma_h$ mirror plane of the {$\{\mathrm{Fe_3}\}$} molecule constrains the dipole moment to lie parallel to the plane of the triangle (i.e. $p_z = 0$). We can express $\Vec{p_i}$ as
\begin{eqnarray}
\Vec{p}_i &=& \Vec{p}_{i1} + \omega \Vec{p}_{i2} + \omega^2\Vec{p}_{i3} \nonumber \\
&=& p_i[(cos\phi+\omega cos(\phi+\beta)+\omega^2cos(\phi+2\beta)\hat{x} \nonumber \\
& & +(sin\phi+\omega sin(\phi+\beta)+\omega^2sin(\phi+2\beta)\hat{y}] \nonumber\\
&=& \frac{3}{2}p_i[(cos\phi - isin\phi)\hat{x}+(sin\phi - icos\phi)\hat{y})] \nonumber\\
&=& \frac{3}{2}[(p_{ix}-ip_{iy})\hat{x}+(p_{iy} + ip_{ix})\hat{y})]
\label{cdip}   
\end{eqnarray}
Here $\beta=120^o$ as shown in Fig.~\ref{dip}. The angle $\phi$ is, in general, non-zero since the magnitude of the spin-projection at each magnetic sites are different. Substituting Eq.~\ref{cdip} in Eq.~\ref{SE1}, we can express the dipole matrix element Eq.~\ref{se} as
\begin{eqnarray}
e\Vec{E}\cdot\Vec{d} &=& e\frac{3}{2}\sum_{i=1}^{(N+1)^2}a_{i1}^*b_{i1}[(p_{ix}-ip_{iy})E_x +(p_{iy} + ip_{ix})E_y)] \nonumber \\
&=& eE\frac{3}{2}\sum_{i=1}^{(N+1)^2}a_{i1}^*b_{i1}[(p_{ix}-ip_{iy}) +(p_{iy} + ip_{ix})]  
\label{me}   
\end{eqnarray}
The second line in Eq.~\ref{me} is obtained by assuming that the electric field is applied in the plane of the triangle such that $E_x = E_y = E$. The strength of the dipole coupling can be  expressed as
\begin{equation}
\abs{\Vec{d}}= \frac{3}{2}\abs{\sum_{i=1}^{(N+1)^2}a_{i1}^*b_{i1}[(p_{ix}+p_{iy}) +i(p_{ix}-p_{iy})]} 
\label{SE2}   
\end{equation}

Therefore, the strength of the spin-electric coupling depends on the dipole moments of all the $(N+1)^2$ inequivalent spin configurations. Since for a given spin configuration, we can calculate the components of the dipole moment from DFT calculations, Eq.~\ref{SE2} provides a way to calculate the coupling strength of any realistic triangular single molecule magnet with half-integer spin.  

In this work we are primarily interested in $s = 5/2$ triangle. Therefore, our work mostly focused on $\{\mathrm{Fe_3}\}$ MM. However, in the next subsection, we briefly discuss $s = 1/2$ and $s = 3/2$ triangular MMs to illustrate how Eq.~\ref{SE2} can be utilized to calculate spin-electric coupling in these MMs.

\subsection*{Spin-electric coupling in $s=1/2$ and $s=3/2$ triangular MMs}

For $s=1/2$ molecules, N = 0 in Eq.~\ref{SE2}. The GS for each $S_z$ consists of three spin configurations, which are related by $C_3$ symmetry. Therefore, only one spin basis state is independent, and the spin-electric coupling is determined by the dipole moment of that state. For a $s=1/2$ triangle, the coefficients $a = b = 1/\sqrt{3}$, and $p=\sqrt{p_x^2 + p_y^2}$. Then the  coupling strength in Eq.~\ref{SE2} reduces to  
\begin{equation}
d = \frac{p}{\sqrt{2}}
\label{SE_Cu3}   
\end{equation}
This is precisely the coupling that we have obtained in our earlier work on $\{\mathrm{Cu_3}\}$ and $\{\mathrm{V_3}\}$ MMs\cite{Islam2010,Nossa2023}.

For $s=3/2$ molecules, N = 1 in Eq.~\ref{SE2}. The chiral ground state of $s=3/2$ triangular MM consists of twelve basis states, of which, only four of them are independent. We solve the Hamiltonian (Eq.~\ref{H_chi}) in the GS manifold. The amplitudes of the basis states of $\Psi^1_{GS}(S_z=1/2, \chi=+1)$ are listed in the Table~\ref{tab_basis1}.  

\begin{table}[h]
\centering
\caption{The complex coefficients of 12 spin basis contained in the $\Psi_{GS}^1$ of $s=3/2$ MM. The row labels  $B_is$ (i=1,2,3,4) corresponds to the subgroup of spin states that are related by $C_3$ symmetry and $a_{ij}s$ (j=1,2,3) are the amplitudes of the corresponding spin states in Eq.~\ref{gs}.}
\begin{tabular}{|c|c|c|c|} \hline
$B_1$     & $\ket{\frac{-3}{2} \frac{3}{2} \frac{1}{2}}$ & $\ket{\frac{1}{2} \frac{-3}{2} \frac{3}{2}}$ & $\ket{\frac{3}{2} \frac{1}{2} \frac{-3}{2}}$   \\  \hline
$a_{1j}$   & -0.1581 + 0.2739i & -0.1581 - 0.2739i &  0.3162 + 0.0000i \\ \hline
$B_2$     & $\ket{\frac{1}{2} \frac{1}{2} \frac{-1}{2}}$  & $\ket{\frac{-1}{2} \frac{1}{2} \frac{1}{2}}$ & $\ket{\frac{1}{2} \frac{-1}{2} \frac{1}{2}}$ \\  \hline
$a_{2j}$   & -0.0000 + 0.1826i & -0.1581 - 0.0913i & 0.1581 - 0.0913i \\ \hline
$B_3$    & $\ket{\frac{1}{2} \frac{3}{2} \frac{-3}{2}}$  & $\ket{\frac{-3}{2} \frac{1}{2} \frac{3}{2}}$ & $\ket{\frac{3}{2} \frac{-3}{2} \frac{1}{2}}$ \\  \hline
$a_{3j}$   & -0.3162 - 0.0000i & 0.1581 - 0.2739i & 0.1581 + 0.2739i \\ \hline
$B_4$     & $\ket{\frac{3}{2} \frac{-1}{2} \frac{-1}{2}}$  & $\ket{\frac{-1}{2} \frac{3}{2} \frac{-1}{2}}$ & $\ket{\frac{-1}{2} \frac{-1}{2} \frac{3}{2}}$ \\  \hline
$a_{4j}$   & -0.2739 - 0.1581i & -0.1581 - 0.2739i & -0.0000 + 0.3162i \\ \hline
\end{tabular}
\label{tab_basis1}
\end{table}
We note that the ratios $a_{i2}/a_{i1}$ and $a_{i3}/a_{i1}$ in each subgroup of bases are $\omega$ and $\omega^2$, respectively. The amplitudes of each basis state in $\Psi^2_{GS}(S_z=1/2, \chi=-1)$ has the same magnitude as listed in the table but the phases are opposite. Since now all the coefficients are known, one can calculate the spin-electric coupling in $s=3/2$ molecules using Eq.~\ref{SE2} simply by calculating dipole moment of the four independent spin states.

\subsection*{Spin-electric coupling in $s=5/2$ $\{\mathrm{Fe_3}\}$ triangular MM}

We now turn our focus on $s=5/2$ triangular MM, namely $\{\mathrm{Fe_3}\}$ SMM. It has three magnetic centers arranged in an equilateral triangle with local spin of each Fe atom is $s=5/2$. The spin manifold of the molecule is much larger compared to the $s=3/2$ manifold, and is given by the irreducible representations of the rotation group of the form $nD^S$ (n is the number of identical irreducible representation $D^S$) as $D^{(15/2)} \bigoplus 2D^{(13/2)} \bigoplus 3D^{(11/2)} \bigoplus 4D^{(9/2)} \bigoplus 5D^{(7/2)} \bigoplus$ $6D^{(5/2)} \bigoplus 4D^{(3/2)} \bigoplus 2D^{(1/2)}$ and has a total of 216 spin states. For the $s=5/2$ triangular MMs, N = 2 in Eq.~\ref{SE2}. The chiral ground state consists of 27 basis states but only 9 of them are independent, and the remaining states are related by $C_3$ symmetry. 

As in the case of $s=3/2$, we have solved the Hamiltonian (Eq.~\ref{H_chi}) in this GS manifold. The amplitudes of the basis states of $\Psi^1_{GS}(S_z=1/2, \chi=+1)$ are listed in the Table~\ref{tab_basis}.

\begin{table}[h]
\centering
\caption{The 27 spin basis contained in the $\Psi_{GS}^1$ of Hamiltonian Eq.~\ref{H_chi} of a $s=5/2$ triangle MM. The row labels  $B_is$ (i=1,2,$\cdots$ 9) corresponds to the subgroup of spin states that are related by $C_3$ symmetry and $a_{ij}s$ (j=1,2,3) are the amplitudes of the corresponding spin states in Eq.~\ref{gs}.}
\begin{tabular}{|c|c|c|c|} \hline
$B_1$     & $\ket{\frac{-5}{2} \frac{3}{2} \frac{3}{2}}$ & $\ket{\frac{3}{2} \frac{-5}{2} \frac{3}{2}}$ & $\ket{\frac{3}{2} \frac{3}{2} \frac{-5}{2}}$   \\  \hline
$a_{1j}$   & 0.1260 - 0.2182i & 0.1260 + 0.2182i &  -0.2520 - 0.0000i \\ \hline
$B_2$     & $\ket{\frac{-3}{2} \frac{-1}{2} \frac{5}{2}}$  & $\ket{\frac{5}{2} \frac{-3}{2} \frac{-1}{2}}$ & $\ket{\frac{-1}{2} \frac{5}{2} \frac{-3}{2}}$ \\  \hline
$a_{2j}$   & 0.0445 - 0.2315i & 0.1782 + 0.1543i & -0.2227 + 0.0772i \\ \hline
$B_3$    & $\ket{\frac{-3}{2} \frac{5}{2} \frac{-1}{2}}$  & $\ket{\frac{-1}{2} \frac{-3}{2} \frac{5}{2}}$ & $\ket{\frac{5}{2} \frac{-1}{2} \frac{-3}{2}}$ \\  \hline
$a_{3j}$   & 0.1782 - 0.1543i & 0.0445 + 0.2315i & -0.2227 - 0.0772i \\ \hline
$B_4$     & $\ket{\frac{-5}{2} \frac{1}{2} \frac{5}{2}}$  & $\ket{\frac{5}{2} \frac{-5}{2} \frac{1}{2}}$ & $\ket{\frac{1}{2} \frac{5}{2} \frac{-5}{2}}$ \\  \hline
$a_{4j}$   & -0.0996 + 0.1725i & -0.0996 - 0.1725i & 0.1992 + 0.0000i \\ \hline
$B_5$    & $\ket{\frac{-5}{2} \frac{5}{2} \frac{1}{2}}$  & $\ket{\frac{1}{2} \frac{-5}{2} \frac{5}{2}}$ & $\ket{\frac{5}{2} \frac{1}{2} \frac{-5}{2}}$ \\  \hline
$a_{5j}$   & -0.0996 + 0.1725i & -0.0996 - 0.1725 & 0.1992 + 0.0000i \\ \hline
$B_6$     & $\ket{\frac{-1}{2} \frac{1}{2} \frac{1}{2}}$  & $\ket{\frac{1}{2} \frac{-1}{2} \frac{1}{2}}$ & $\ket{\frac{1}{2} \frac{1}{2} \frac{-1}{2}}$ \\  \hline
$a_{6j}$   & 0.0797 - 0.1380i & 0.0797 + 0.1380i & -0.1594 - 0.0000i \\ \hline
$B_7$    & $\ket{\frac{-3}{2} \frac{1}{2} \frac{3}{2}}$  & $\ket{\frac{3}{2} \frac{-3}{2} \frac{1}{2}}$ & $\ket{\frac{1}{2} \frac{3}{2} \frac{-3}{2}}$ \\  \hline
$a_{7j}$   & 0.0398 + 0.1380i & -0.1394 - 0.0345i & 0.0996 - 0.1035i \\ \hline
$B_8$    & $\ket{\frac{-3}{2} \frac{3}{2} \frac{1}{2}}$  & $\ket{\frac{1}{2} \frac{-3}{2} \frac{3}{2}}$ & $\ket{\frac{3}{2} \frac{1}{2} \frac{-3}{2}}$ \\  \hline
$a_{8j}$   &  -0.1394 + 0.0345i &  0.0398 - 0.1380i & 0.0996 + 0.1035i \\ \hline
$B_9$    & $\ket{\frac{-1}{2} \frac{-1}{2} \frac{3}{2}}$  & $\ket{\frac{3}{2} \frac{-1}{2} \frac{-1}{2}}$ & $\ket{\frac{-1}{2} \frac{3}{2} \frac{-1}{2}}$ \\  \hline
$a_{9j}$   &  -0.1127 - 0.0000i & 0.0563 - 0.0976i & 0.0563 + 0.0976i \\ \hline
\end{tabular}
\label{tab_basis}
\end{table}

\begin{figure}[ht]
\centering{\includegraphics[width=0.53\textwidth]{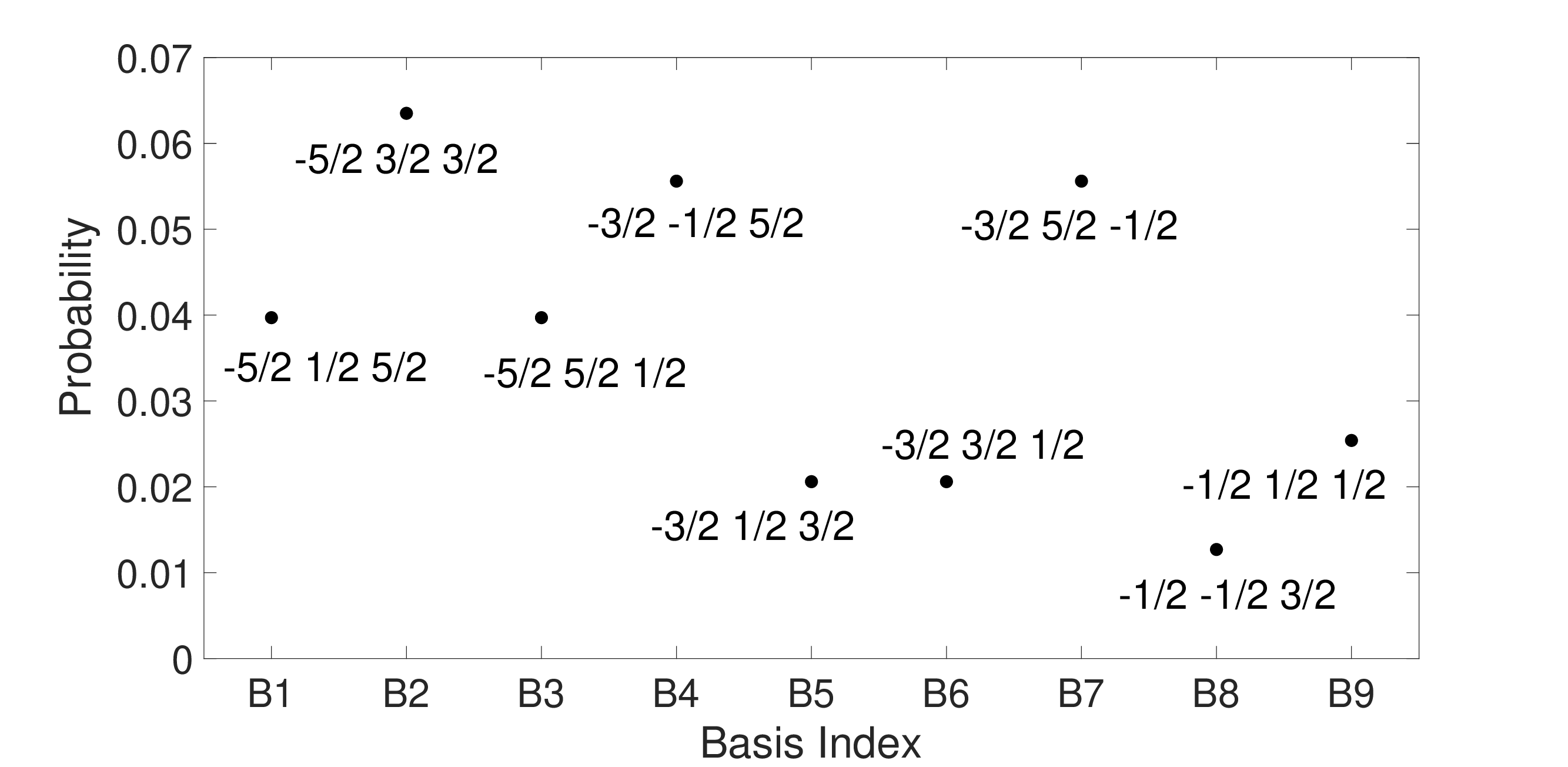}}
\caption{Probability of the nine inequivalent spin configurations in the GS.}
\label{Prob}
\end{figure}

In Fig.~\ref{Prob}, we have plotted the probability of the nine inequivalent basis states in the GS of the molecule. We note from the figure that the state, $\ket{\frac{-5}{2} \frac{3}{2} \frac{3}{2}}$, has the highest probability and, therefore, is the basis state with the lowest energy in the GS. It is also evident that there are three pair of states that are degenerate. Therefore, the GS contains basis states with six different energies.  

\section{Computational methods}
\subsection*{Implementation of non-collinear magnetism in NRLMOL DFT code}
\label{ncm}

\begin{figure}[ht]
\centering{\includegraphics[width=0.35\textwidth]{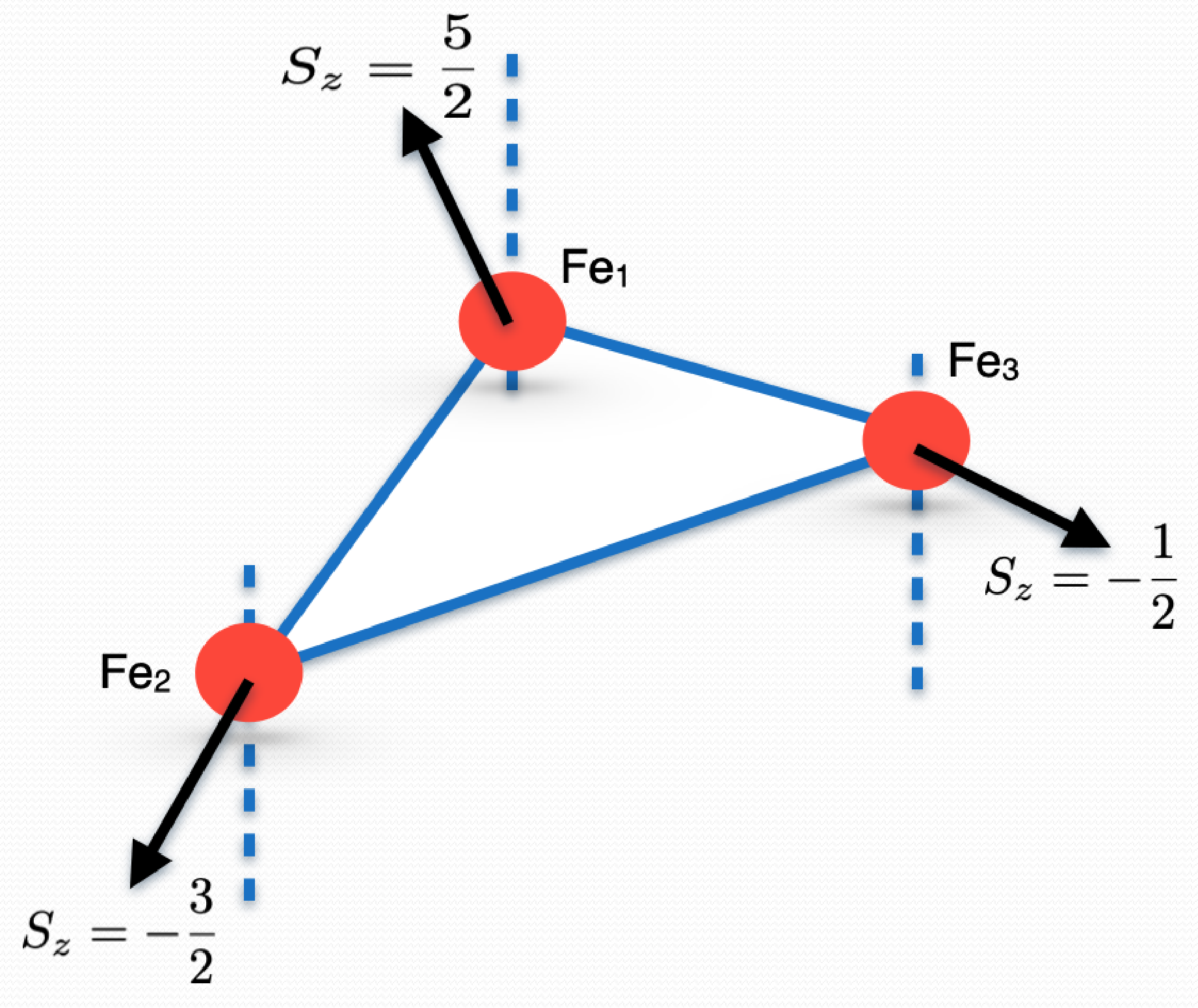}}
\caption{A visual representation of one of the spin basis $\ket{\frac{5}{2} \frac{-3}{2} \frac{-1}{2}}$ involved in the GS of $\{\mathrm{Fe_3}\}$ MM. Note that the local spin of each Fe atom is 5/2 but the spin projection at different sites are different, resulting in NCL magnetism.}
\label{spin_st}
\end{figure}

In Sec.~\ref{SEcoupling} we have discussed the theoretical method for calculating spin-electric coupling in $s=5/2$ triangular molecule. The calculation of the coupling strength requires one to calculate electric dipole moments of 9 different NCL spin configurations. In Fig.~\ref{spin_st}, we have plotted the schematic of one of the spin configurations, $\ket{\frac{5}{2} \frac{-3}{2} \frac{-1}{2}}$ in the GS. The state is labelled by the spin-projections at the respective sites of the  $\{\mathrm{Fe_3}\}$ triangle as shown in the figure. 

The previous version of the DFT code NRLMOL\cite{Pederson1990,Jackson1990,Porezag1999} that is used for electronic structure calculations works only for collinear magnetic molecules. Therefore, to calculate spin-electric coupling in $\{\mathrm{Fe_3}\}$ molecule, we have first implemented NCL magnetism in NRLMOL following the procedure described by K\"{u}bler {\it et. al.} \cite{Kubler1988,Kubler2_1988}. Note that this approach has some similarity with a theoretical formalism that we introduced a few years ago to describe the many-electron GS spin multiplet wave functions of Mn$_{12}$ molecular magnet by means of DFT methods\cite{Michalak2010}. An overview of NCL magnetism and its implementation in NRLMOL is described below.  

In Kohn-Sham (KS) DFT, the effective potential within the local spin density approximation (LSDA)\cite{Parr1994} is the sum of three contributions, 
\begin{equation}
V_\mathrm{eff}=V_\mathrm{coul}[\rho_\mathrm{tot}]+V_\mathrm{ext}[\rho_\mathrm{tot}]+V_\mathrm{XC} [\rho_{\uparrow},\rho_{\downarrow}].
\label{pot}
\end{equation}
The Coulomb potential $V_{coul}$ and the external potential $V_{ext}$ do not depend on the spin, so they can be calculated using the total density $\rho_{tot}$. The exchange-correlation (XC) potential in standard DFT is well-defined only for collinear magnetism and is a functional of pure spin-up density $\rho_{\uparrow}$ and spin-down density $\rho_{\downarrow}$.  

\begin{figure}[ht]
\centering{\includegraphics[width=0.47\textwidth]{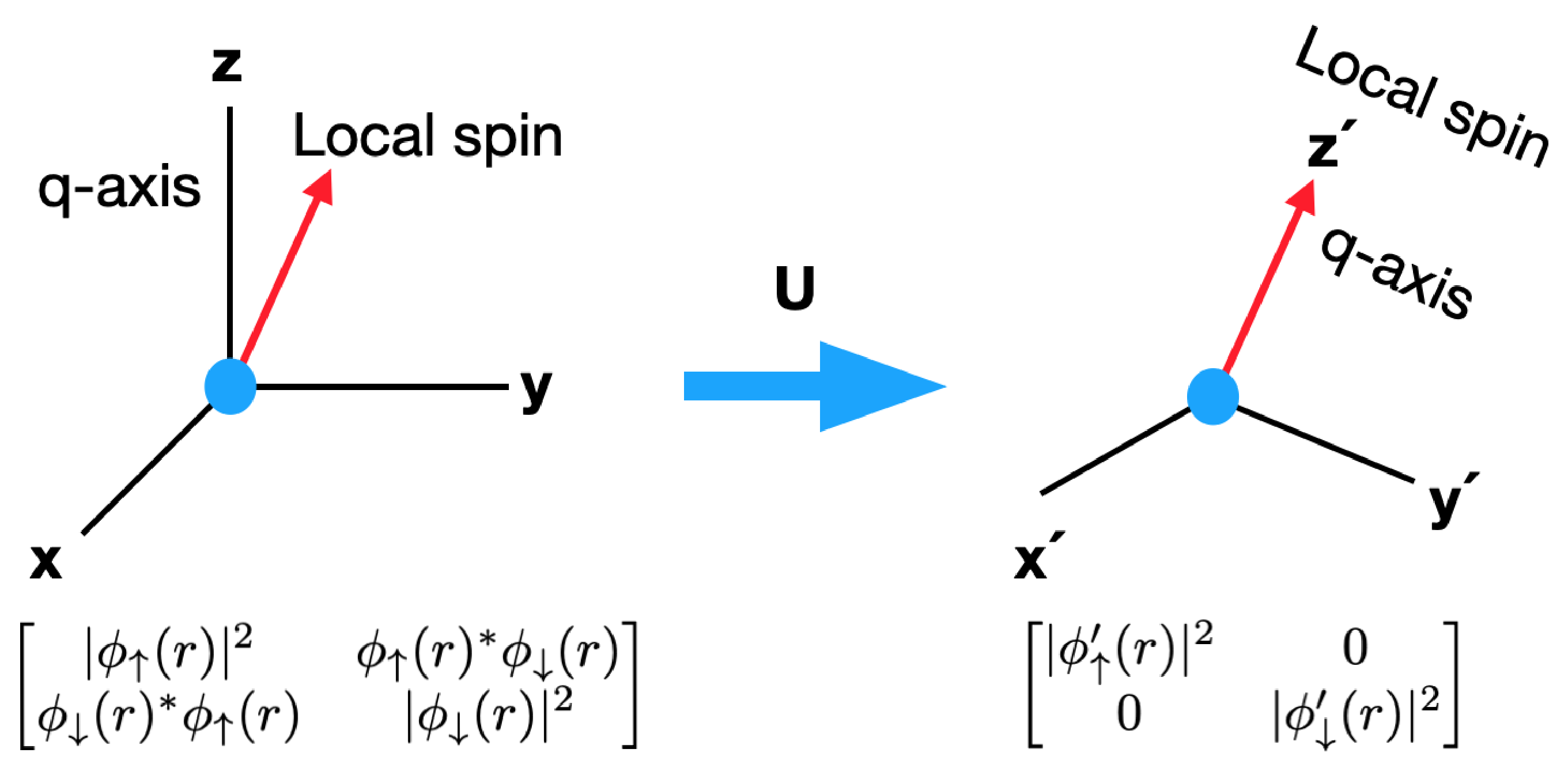}}
\caption{On the left a general NCL density matrix at a point (blue dot) expressed with respect to the global q-axis, which is transformed to a diagonal density matrix (on the right) with respect to a local q-axis by the unitary transformation U. This local diagonal density matrix is used to calculate XC potential of a NCL magnetic system at that point.}
\label{Rdmat}
\end{figure}

For a NCL magnetic systems such as frustrated triangular MM, the single particle states are mixed states of spin-up and spin-down states. Such single particle states are described by spinors and the corresponding density matrices are 2 $\times$ 2 matrices with both diagonal and off-diagonal elements as shown in Eq.~\ref{dmat}. Note that the spin-orbit coupling also mixes the spin-up and spin-dn states but in this case non-collinearity arises from geometric frustration. 

\begin{eqnarray}
\psi_i(r) &=& \begin{bmatrix} \phi_{i\uparrow}(r) \\ \phi_{i\downarrow}(r)\end {bmatrix} \nonumber \\
\rho_i(r)&=& \begin{bmatrix} \abs{\phi_{i\uparrow}(r)}^2 & \phi_{i\uparrow}^*(r)\phi_{i\downarrow}(r) \\ \phi_{i\downarrow}^*(r)\phi_{i\uparrow}(r) & \abs{\phi_{i\downarrow}(r)}^2 \end {bmatrix} 
\label{dmat}   
\end{eqnarray}
This generalized density matrix must be used to calculate the potential at each point $\Vec{r}$ specified by a predetermined mesh to solve the KS equation. Therefore, for NCL DFT, the calculation of XC potential $V_{xc}$ requires spacial considerations. In order to utilize standard XC potential to calculate XC potential at a point for a NCL magnetic system, we first perform a unitary transformation that diagonalizes the density matrix at that point which is schematically described in Fig~\ref{Rdmat}. The eigenvalues of the density matrix correspond to the spin-up and spin-dn densities in rotated coordinates in which the quantization axis (q-axis) coincides with the local spin direction at that particular point of the mesh. We can now use $|\phi'_{\uparrow}(r)|^2$ and $|\phi'_{\downarrow}(r)|^2$ to calculate XC potential at $\Vec{r}$ using standard collinear XC potential. To add this XC potential with other potentials in Eq.~\ref{pot}, we must rotate it back in the global quantization direction, which is performed by applying the inverse transformation $U^{-1}$.  

We have implemented this procedure in NRLMOL, both for LSDA and Perdew-Burke-Ernzerhof generalized gradient approximation (PBE-GGA)\cite{Perdew1996}. Note that for GGA, potentials are functional of both density and gradient of density. The gradient of density is evaluated from $|\phi'_{\uparrow}(r)|^2$ and $|\phi'_{\downarrow}(r)|^2$ in the rotated coordinates. 

While a collinear spin system maybe described by a global q-axis (typically along the z-axis of the coordinate system) for all magnetic atoms, in a NCL system, q-axis of each magnetic atom is described locally. In NRLMOL, a NCL spin configuration, for example, the one shown in Fig~\ref{spin_st}, is realized by applying magnetic atom-centered bias magnetic fields of the form: 
\begin{equation}
\bm{B}_\mathrm{loc}=\bm{B}_0e^{\alpha(\bm{r}-\bm{R}_A)^2}
\label{qaxis}
\end{equation}
along a direction determined by the spin configurations. Here, $B_0$ and $\alpha$ are the parameters to control the strength and the extent of the local magnetic field centered at the atom at $\bm{R}_A$. For $\{\mathrm{Fe_3}\}$ the bias potential is applied only on the three Fe atoms in the molecule. The exponential decay of the bias field ensures the potential confined withing the atomic sphere of the atoms.

In order to obtain a smooth convergence of a specific spin configuration within the ground state, the DFT calculations are performed in three steps. In the first step, we perform a collinear spin-polarized calculation of the $\{\mathrm{Fe_3}\}$ molecule. The converged density is then used as the starting density for all NCL calculations. In the second step, we have add a bias fields to realize a specific spin-configuration. It requires several iterations to stabilize local spin orientations of the Fe atoms. Then, in the third step, we remove the bias field to achieve self-consistent convergence of a specific spin configuration. 

A final remark on the computational aspect of this work. In the third step, as described above, when we remove the local bias field, the spins maintain the desired spin configuration in all cases except for $\ket{\frac{-1}{2} \frac{1}{2} \frac{1}{2}}$. For this configuration we have used dipole moment in presence of the local bias field which gives slightly larger $p_y$ component but does not have significant effect on the dipole coupling of the molecule. Furthermore, in order to realize different NCL spin configurations within the ground state manifold, we have relaxed all the symmetry constraint of the $D_{3h}$ point group. This introduces small numerical difference for different degenerate states. Despite this small numerical issues, we believe our calculations provide a good estimate of the spin-electric coupling in $\{\mathrm{Fe_3}\}$ molecule.

\section{Results and Discussion}
\label{Results}

\subsection{Electronic Structure of $\{\mathrm{Fe_3}\}$ MM}
\label{ElecStruct}

\begin{figure}[ht]
\centering{\includegraphics[width=0.45\textwidth]{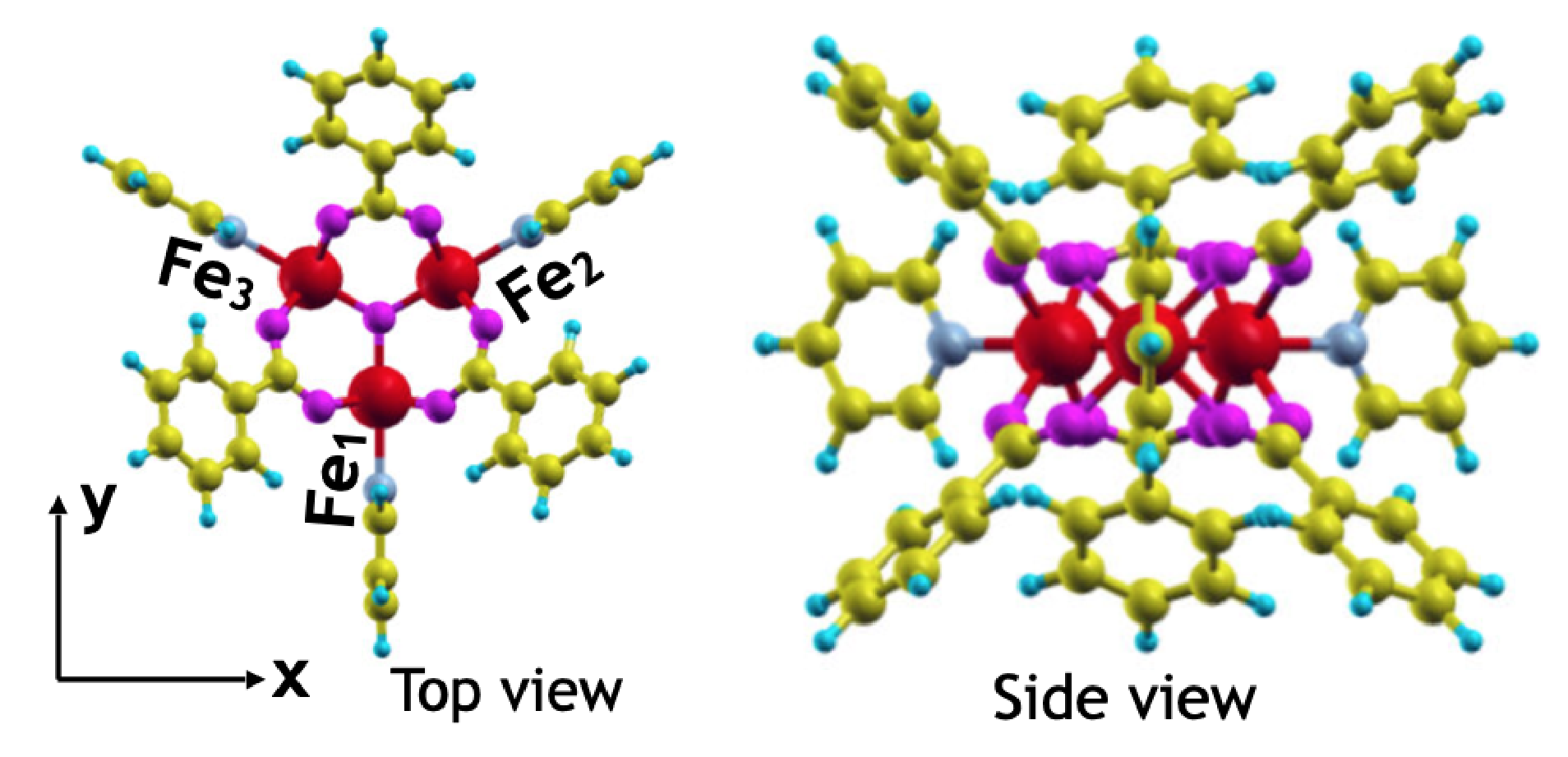}}
\caption{Structure of $\{\mathrm{Fe_3}\}$ molecule showing the three magnetic centers (red) forming equilateral triangle in the x-y plane.}
\label{struct}
\end{figure}

The molecular structure of $\{\mathrm{Fe_3}\}$ MM that is used to study spin-electric coupling in this work is shown in Fig.~\ref{struct}. It belongs to $\mathrm{D_{3h}}$ point group symmetry, which contains twelve symmetry operations including $\mathrm{C_3}$ rotation and $\mathrm{\sigma_h}$ mirror plane. These two symmetries play important role in spin-electric coupling in this molecule. The $\mathrm{C_3}$ results in chiral GS where as $\mathrm{\sigma_h}$ constrains the dipole moment of different spin configurations in the GS to lie in the plane of the triangle. Previously we have studied electronic properties of this molecule using collinear DFT\cite{Johnson2019}.

Our calculations show that the GS spin configuration of the molecule is anti-ferromagnetic with local spin of each Fe atom, $s=5/2$. However, our calculation also shows that there exists a nearly degenerate spin configuration with local spin $s=1/2$ that differ only by about 8.2 meV. Therefore, this molecule is also a good candidate for a spin-crossover MM. Our calculations further show that the molecule has easy plane magnetic anisotropy of 8 Kelvin in its GS spin configuration ($s=5/2$), whereas for $s=1/2$ spin configuration it has easy axis magnetic anisotropy of 3 K. 

\subsection{DFT calculations of spin-electric coupling in $\{\mathrm{Fe_3}\}$ MM}
\label{sec}

Since the spin-electric coupling in frustrated triangular SMMs originates from the spin-induced electric dipole moment resulting from charge redistribution, we have calculated dipole moments of all nine independent spin-configurations in the ground state of $\{\mathrm{Fe_3}\}$ molecule using our newly developed NCL DFT code. The results are shown in Table~\ref{spin_dipole}. Because of $\sigma_h$ symmetry in $\{\mathrm{Fe_3}\}$, dipole moments lie on the x-y plane, and, consequently, $p_z=0$ (since we didn't impose any symmetry explicitly, our numerical calculations show a small $p_z$ component of order $10^{-6}$, which is negligible compared to the in-plane components) .    

\begin{table}[h]
\centering
\caption{Dipole moments of the nine inequivalent spin configurations in units of $ea_0$
where e is the electronic charge and $a_0$ is the Bohr radius. The coefficients in the third 
column are obtained from Table~\ref{tab_basis}.}
\begin{tabular}{|c|c|c|c|c|} \hline
Spin        &   Inequivalent                   & $a_i^*b_i $& $p_x$    & $p_y$    \\  
group       &   spin config.                   &           & ($ea_0$) & ($ea_0$) \\  \hline 
B1 & $\ket{\frac{-5}{2} \frac{3}{2} \frac{3}{2}}$  & 0.0635    &  0.001  & 0.070   \\  \hline    
B2 & $\ket{\frac{-3}{2} \frac{-1}{2} \frac{5}{2}}$ & 0.0556    & -0.038  & 0.011   \\  \hline    
B3 & $\ket{\frac{-3}{2} \frac{5}{2} \frac{-1}{2}}$ & 0.0556    &  0.034  & 0.012   \\  \hline    
B4 & $\ket{\frac{-5}{2} \frac{1}{2} \frac{5}{2}}$  & 0.0397    & -0.080  & 0.088   \\  \hline    
B5 & $\ket{\frac{-5}{2} \frac{5}{2} \frac{1}{2}}$  & 0.0397    &  0.086  & 0.081   \\  \hline    
B6 & $\ket{\frac{-1}{2} \frac{1}{2} \frac{1}{2}}$  & 0.0254    &  0.003  & 0.026   \\  \hline    
B7 & $\ket{\frac{-3}{2} \frac{1}{2} \frac{3}{2}}$  & 0.0206    & -0.020  & 0.030   \\  \hline    
B8 & $\ket{\frac{-3}{2} \frac{3}{2}  \frac{1}{2}}$ & 0.0206    &  0.021  & 0.031   \\  \hline    
B9 & $\ket{\frac{-1}{2} \frac{-1}{2} \frac{3}{2}}$ & 0.0127    & -0.019  &-0.005  \\  \hline   
\end{tabular}
\label{spin_dipole}
\end{table}

\subsection*{Spin-induced electric dipole moment for noncollinear spins}

\begin{figure}[ht]
\centering{\includegraphics[width=0.45\textwidth]{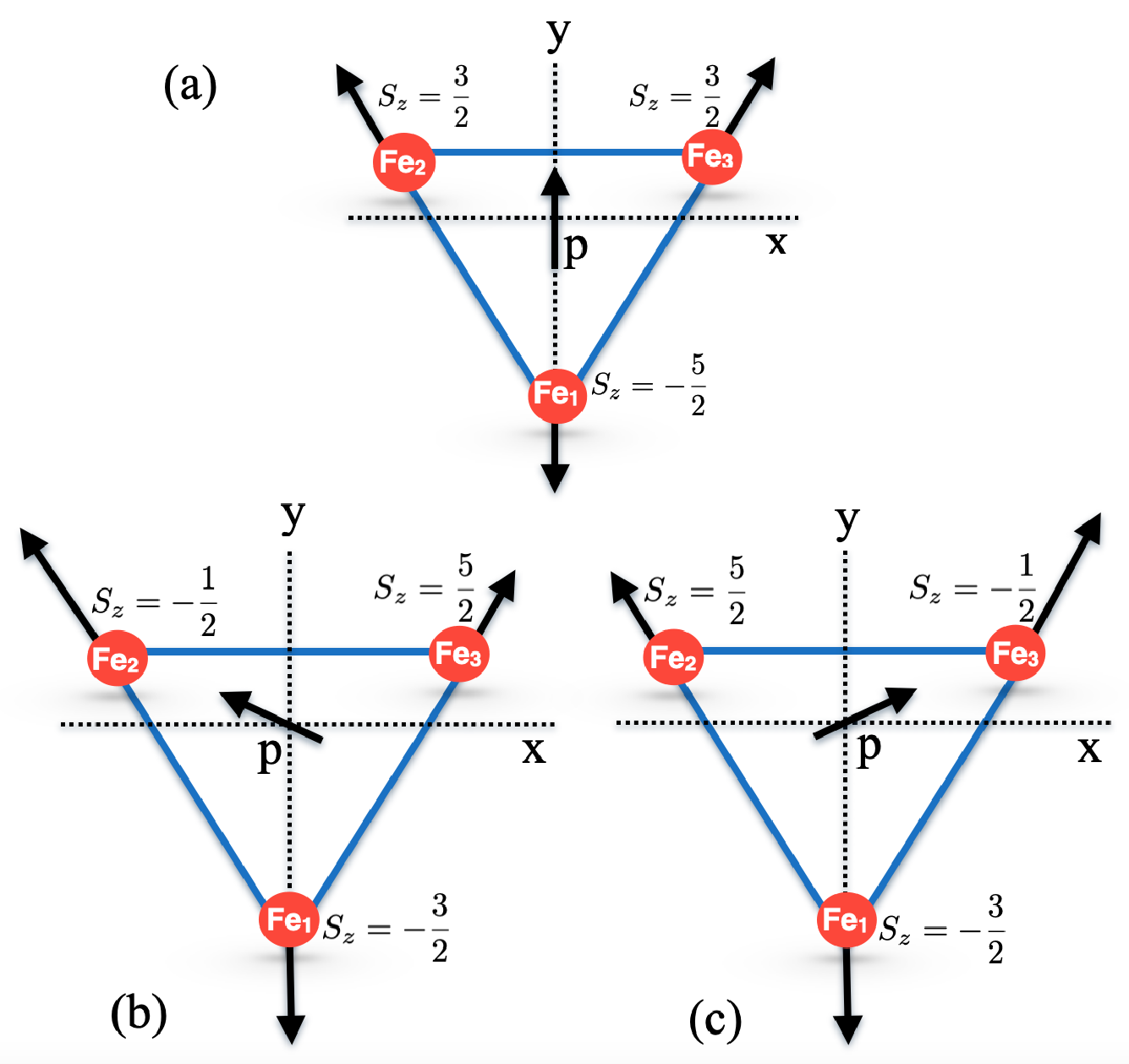}}
\caption{Spin projections of local spins of Fe atoms on x-y plane for a) $\ket{\frac{-5}{2} \frac{3}{2} \frac{3}{2}}$, b) $\ket{\frac{-3}{2} \frac{-1}{2} \frac{5}{2}}$, and c) $\ket{\frac{-3}{2} \frac{5}{2} \frac{-1}{2}}$ spin configurations. The size of the arrow at each Fe site approximately represents the magnitude of the projection. The $S_z=5/2$ projection is the shortest since its projection on x-y plane is smaller compared to the other two $S_z$'s. The arrow at the center of each $\mathrm{Fe_3}$ triangle shows the direction of the spin-induced dipole moments of the corresponding spin configuration.}
\label{ncl_dipole}
\end{figure}

In all our prior works on spin-electric coupling we have used spin 1/2 frustrated triangular SMMs. To calculate coupling strength in the ground state of those molecules, we need to consider only one spin configuration where the spin projection of one of the three magnetic atoms in the triangle is $s_z =-1/2$, and for the other two atoms, projections are $s_z = 1/2$. However, for $\{\mathrm{Fe_3}\}$, the spin projection at different magnetic sites are, in general, different. As discussed in Sec.~\ref{ncm}, realization of such spin-configuration requires NCL arrangement of the local spin of each Fe atoms. In this subsection we discuss how the spin-induced dipole moment depends on different NCL spin configurations. Fig.~\ref{ncl_dipole} shows change of orientation of the dipole moments for the first three spin-configurations in Table~\ref{spin_dipole}.

For the $\ket{\frac{-5}{2} \frac{3}{2} \frac{3}{2}}$ configuration, the dipole points along +y, which is expected since the local spin-projections at the Fe2 and Fe3 sites are the same and, therefore, expected to have the same charge redistribution at these sites. The other two configurations in Fig.~\ref{ncl_dipole}, $\ket{\frac{-3}{2} \frac{-1}{2} \frac{5}{2}}$ and $\ket{\frac{-3}{2} \frac{5}{2} \frac{-1}{2}}$, are related by the interchange of local spins between Fe2 and Fe3 sites. Consequently, the dipole moment has the same magnitude but orientation changes from pointing along Fe2 to Fe3, respectively, and their x-components cancel out. This is also evident for other degenerate pairs in Table~\ref{spin_dipole}, namely (B4, B5) and (B7, B8).

The dependence of spin-induced dipole moment on spin configurations are also evident from Table~\ref{spin_dipole}. We note from the table that more a spin configuration deviates from the collinear configuration, smaller is the dipole moment. For example, $\ket{\frac{-5}{2} \frac{1}{2} \frac{5}{2}}$ has larger dipole moment than $\ket{\frac{-1}{2} \frac{1}{2} \frac{1}{2}}$ ($s_z=1/2$ of a $s = 5/2$ of Fe atoms implies larger polar angle with z-axis). To illustrate better, we have calculated dipole moment of $\ket{\frac{-5}{2} \frac{5}{2} \frac{5}{2}}$ configuration. Note that this configuration is not a basis state of the ground state (\ref{cgs1}) or (\ref{cgs2})  but closest to the classical collinear state. The dipole moment is calculated to be 0.147 $ea_0$, which is larger than any spin configuration in Table~\ref{spin_dipole}. The dependence of dipole moment on spin configuration can be understood intuitively as follows. In the contest of a Hubbard model description, in a frustrated system displaying NCL (and non-planar) classical spin configurations, the electron spins are effectively coupled to the electron charge fluctuation, in a way somewhat mimicking an effective spin-orbit coupling. The quantum ground state of $\mathrm{Fe_3}$ consists of nine independent spin configurations, each with different energy expectation value and a different degree of non-collinearity. A larger NCL angle implies that bonds between Fe atom and its ligands are more stretched, which reduces the charge redistribution within the $Fe_3$ triangle. Hence, the reduced dipole moment.

\subsection{Spin-electric coupling strength in $\{\mathrm{Fe_3}\}$ MM}
\label{cs}

Now we present the main result of this work. We have calculated the coupling strength from Eq.~\ref{SE2} using the data obtained in Table~\ref{spin_dipole}. Our calculation shows that the strength of spin-electric coupling in $\{\mathrm{Fe_3}\}$ MM is 0.030 $ea_0$. It is slightly smaller than that obtained for $\{\mathrm{V_3}\}$ MM (0.035 $ea_0$) but one order of magnitude larger than $\{\mathrm{V_{15}}\}$ MM\cite{Nossa2023}. Note that the strength is an order of magnitude smaller compared to the dipole moment of water, which is about 0.736 $ea_0$.  

One important conclusion of our work is that the spin-electric coupling does not strongly depend on the local spin moment of the magnetic atoms of the frustrated molecule. The reason is that for larger spins, the quantum ground state contains many spin configurations with different dipole moments with different phases. The resulting interference effectively reduces the dipole coupling of the triangular MM with larger local spins. In this sense, $\{\mathrm{V_3}\}$ MM is more effective since only one spin configuration determines the coupling strength.

\section{Summary and conclusions}
\label{Conclusion}

In this work we have studied the spin-electric coupling in frustrated triangular MM. We have developed a general theoretical approach for calculating coupling strength of any half-integer spin triangular magnet. The method involves constructing chiral GSs of these systems by solving Heisenberg model in spin bases of the $s= (2N+ 1)/2$ molecules, with $N = 0, 1, \cdots$. We have then shown that the coupling strength depends only on the dipole moments of different spin configurations involved in the GS, which can be calculated using first-principles DFT method. In this work we have applied the  method to calculate the spin-electric coupling in $\{\mathrm{Fe_3}\}$ MM. Our work shows that in $\{\mathrm{Fe_3}\}$ the coupling strength is of the same order of magnitude as in $\{\mathrm{V_3}\}$, which is a bit surprising consider the large local spin moment of Fe atoms. But we have shown that a triangular magnet with a larger spin also leads to interference, which is detrimental for coupling.   

Since spin frustration in a system leads to Non-collinearity in spins, studying spin-electric coupling in any frustrated system requires NCL magnetism. One of the very important developments of this work is to implement a tool to computationally realize NCL spin states within the framework of NRLMOL DFT code for molecules/clusters. This updated version of the code is expected to play a significant role in studying Dzyaloshinskii-Moriya (DM) interaction\cite{Dzyaloshinsky1958,Moriya_prl_1960,Moriya1960}, which is an important aspect of NCL spin systems, and plays a crucial role in zero-field splitting in frustrated molecular magnets. 

In this work we have focused on the theoretical method and development of computational tools to calculate spin-electric coupling in $\{\mathrm{Fe_3}\}$ MM. However, a direct comparison with experimental observation require further development. Firstly, to observe the transition between the states of opposite chirality requires one to calculate zero-field splitting between these states due to DM interaction. Secondly, the experimental observation of spin-electric coupling was performed by measuring the change in the EPR spectra when an in-plane electric field is introduced. The DFT calculations of zero-field splitting and spectra are very challenging tasks, particularly for a complex spin system like $\{\mathrm{Fe_3}\}$ MM. We plan to address these challenges in our future work.

\appendix*

\section*{Acknowledgment}
\label{sec:Acknowledgment} 
MFI and MRP were supported as part of the Molecular Magnetic Quantum Materials EFRC, an Energy Frontier Research Center funded by the U.S. Department of Energy, Office of Science, Office of Basic Energy Sciences under Award Number DE-SC0019330. KW's contribution to this work were supported  by the CCS FLOSIC project under the U.S. Department of Energy, Office of Science, Office of Basic Energy Sciences, under award number DE-SC0018331.
Work performed at LNU was supported by the School of Computer Science, Physics and Mathematics at Linnaeus University, the Swedish Research Council under Grants No:
621-2010-5119 and 621-2014-4785, by the Carl
Tryggers Stiftelse through Grant No. CTS 14:178 and the NordForsk research network 080134 ``Nanospintronics: theory and simulations". Computational
resources for early calculations have been provided by the
Lunarc Center for Scientific and Technical Computing at Lund
University. Final calculations were performed on the Jakar cluster at UTEP.

\bibliography{Fe3}

\end{document}